\pdfoutput=1
\RequirePackage{ifpdf}
\ifpdf 
\documentclass[pdftex]{sigma}
\else
\documentclass{sigma}
\fi

\usepackage{mathrsfs}
\usepackage{braket}		
\usepackage{slashed}
\usepackage{twistor}
\usepackage[all]{xy}
\usepackage{tikz-cd}

\newcommand{\ba}{\mathbf{a}}

\newcommand{\h}{{\mathbf h}}

\renewcommand{\d}{\mathrm{d}}

\newcommand{\eps}{\epsilon}

\newcommand{\nbar}{\bar{\nabla}}
\newcommand{\bigma}{\boldsymbol{\sigma}}

\numberwithin{equation}{section}

\newtheorem{Theorem}{Theorem}[section]
\newtheorem*{Theorem*}{Theorem}

 { \theoremstyle{definition}

 }

\begin{document}
\allowdisplaybreaks

\newcommand{\arXivNumber}{2110.06066}

\renewcommand{\thefootnote}{}

\renewcommand{\PaperNumber}{016}

\FirstPageHeading

\ShortArticleName{Celestial $w_{1+\infty}$ Symmetries from Twistor Space}

\ArticleName{Celestial $\boldsymbol{w_{1+\infty}}$ Symmetries from Twistor Space\footnote{This paper is a~contribution to the Special Issue on Twistors from Geometry to Physics in honor of Roger Penrose. The~full collection is available at \href{https://www.emis.de/journals/SIGMA/Penrose.html}{https://www.emis.de/journals/SIGMA/Penrose.html}}}

\Author{Tim ADAMO~$^{\rm a}$, Lionel MASON~$^{\rm b}$ and Atul SHARMA~$^{\rm b}$}

\AuthorNameForHeading{T.~Adamo, L.~Mason and A.~Sharma}

\Address{$^{\rm a)}$~School of Mathematics and Maxwell Institute for Mathematical Sciences, \\
\hphantom{$^{\rm a)}$}~University of Edinburgh, EH9 3FD, UK}
\EmailD{\href{mailto:t.adamo@ed.ac.uk}{t.adamo@ed.ac.uk}}

\Address{$^{\rm b)}$~The Mathematical Institute, University of Oxford, OX2 6GG, UK}
\EmailD{\href{mailto:lmason@maths.ox.ac.uk}{lmason@maths.ox.ac.uk}, \href{mailto:atul.sharma@maths.ox.ac.uk}{atul.sharma@maths.ox.ac.uk}}

\ArticleDates{Received November 22, 2021, in final form February 17, 2022; Published online March 08, 2022}

\Abstract{We explain how twistor theory represents the self-dual sector of four dimensional gravity in terms of the loop group of Poisson diffeomorphisms of the plane via Penrose's non-linear graviton construction. The symmetries of the self-dual sector are generated by the corresponding loop algebra $Lw_{1+\infty}$ of the algebra $w_{1+\infty}$ of these Poisson diffeomorphisms. We show that these coincide with the infinite tower of soft graviton symmetries in tree-level perturbative gravity recently discovered in the context of celestial amplitudes. We use a~twistor sigma model for the self-dual sector which describes maps from the Riemann sphere to the asymptotic twistor space defined from characteristic data at null infinity $\scri$. We show that the OPE of the sigma model naturally encodes the Poisson structure on twistor space and gives rise to the celestial realization of $Lw_{1+\infty}$. The vertex operators representing soft gravitons in our model act as currents generating the wedge algebra of $w_{1+\infty}$ and produce the expected celestial OPE with hard gravitons of both helicities. We also discuss how the two copies of $Lw_{1+\infty}$, one for each of the self-dual and anti-self-dual sectors, are represented in the OPEs of vertex operators of the 4d ambitwistor string.}


\Classification{83C60; 81U20; 32L25}

\begin{flushright}
\begin{minipage}{83mm}
\it
Dedicated to our friend and mentor Roger Penrose\\ on the occasion of his 90th birthday and the recent\\ award of his Nobel prize in physics.
\end{minipage}
\end{flushright}

\renewcommand{\thefootnote}{\arabic{footnote}}
\setcounter{footnote}{0}

\section{Introduction}
Among other things, Roger Penrose is famous in general relativity for his introduction of null infinity, $\scri$, as the geometry underpinning the asymptotics of massless space-time radiation fields~\cite{Penrose:1962ij,Penrose:1965am}.
In recent years, there has been a resurgence in the study of asymptotic symmetries and scattering amplitudes at null infinity $\scri$, much of it aimed at formulating a notion of holography for asymptotically flat space-times (cf.~\cite{Aneesh:2021uzk,Pasterski:2021rjz,Raclariu:2021zjz, Strominger:2017zoo} for recent reviews). In fact, the notion of reconstructing `bulk' space-times and their physics holographically at $\scri$ dates back to the 1970s and the work of Newman and Penrose~\cite{Newman:1976gc,Penrose:1976js,Penrose:1976jq}. One of the main outputs of this work was the \emph{non-linear graviton} construction, where (complex) space-times with self-dual curvature arise from deformations of the complex structure on twistor spaces. When these are `asymptotic' twistor spaces, the non-linear graviton is intrinsically holographic, as the deformed complex structure is constructed directly from the (complexified) characteristic data (i.e., the self-dual asymptotic shear) of an asymptotically flat, radiative self-dual space-time at $\scri$~\cite{Eastwood:1982}.

Much of the recent work on `celestial holography' has focused on the interplay between asymptotic symmetries and soft particles~\cite{Strominger:2013lka, Strominger:2013jfa}. For example, at leading order in the soft momentum, soft gravitons are related to BMS supertranslations via a Ward identity~\cite{He:2014laa}; there are now many generalizations to subleading orders and other theories (cf.~\cite{Strominger:2017zoo} and references therein) which can also be understood in terms of an interplay between asymptotic symmetries and twistor or ambitwistor data~\cite{Adamo:2015fwa, Adamo:2014yya,Geyer:2014lca}. By expressing scattering amplitudes in terms of a conformal primary basis on the celestial sphere~\cite{Pasterski:2017kqt, Pasterski:2016qvg}, it is clear that there is actually an \emph{infinite} tower of conformal soft graviton theorems arising when the soft external graviton has scaling dimension $\Delta=2,1,0,-1,\ldots$~\cite{Adamo:2019ipt,Donnay:2018neh,Guevara:2019ypd,Puhm:2019zbl}. For a positive helicity soft graviton, this infinite tower of soft theorems can be organized into the algebra $w_{1+\infty}$ (or more precisely, the loop algebra of the wedge algebra of $w_{1+\infty}$)~\cite{Guevara:2021abz,Himwich:2021dau,Jiang:2021ovh,Strominger:2021lvk}.

It has long been known that the algebra $w_{1+\infty}$ classically describes canonical transformations of a plane~\cite{Bakas:1989xu, Hoppe:1988gk}. Over the years, a number of authors have linked this to self-dual gravity via the non-linear graviton construction~\cite{Boyer:1985aj,Mason:1990,Park:1989fz,Park:1989vq} of deformed twistor spaces for self-dual space-times. The deformed twistor spaces are glued together by patching functions that can be expressed as maps from a neighbourhood of the equator of the Riemann sphere to canonical transformations of the 2-dimensional fibres of the twistor space over this sphere, as explained by Penrose himself in his original paper~\cite{Penrose:1976js}. Thus, the Lie algebra, $Lw_{1+\infty}$, of the loop group of canonical transformations acts on this space of patching functions for twistor space and hence on the space of all self-dual Ricci-flat metrics. Although $Lw_{1+\infty}$ transformations act by diffeomorphisms and hence resemble gauge transformations, generically they are not global and have singularities. They define genuine deformations of the twistor space and are not, strictly speaking, symmetries. Such constructions making use of singular gauge transformations on twistor space to transform one solution to another are standard in twistor formulations of classical B\"acklund transformations in the study of integrable systems (cf.~\cite{Mason:1988xd,Mason:1988ea,Mason:1991rf,Woodhouse:1987,Mason:1988ernst}). These ideas were developed into a twistor formulation of the $Lw_{1+\infty}$ symmetries via a recursion operator based on such a Backlund transformation to generate the loop algebra from coordinate symmetries in \cite{Dunajski:2000iq}.\looseness=-1

In the non-linear graviton construction, the self-dual space-time is recovered as the four-dimensional family of rational holomorphic curves in twistor space of degree one. Recently, we introduced sigma models in twistor space for such holomorphic curves~\cite{Adamo:2021bej} whose on-shell action is equal to the K\"ahler scalar (or first Pleba\'nski scalar~\cite{Plebanski:1975wn}) of the associated self-dual space-time. These `twistor sigma models' can be used to construct gravitational MHV scattering amplitudes directly from general relativity, and at higher-degree build the full tree-level $S$-matrix of gravity via a natural family of generating functionals. In this paper, we show how the loop algebra of~$w_{1+\infty}$ and the infinite tower of soft graviton theorems is realised in terms of these twistor sigma models. We will also see that the action of~$Lw_{1+\infty}$ can be lifted to 4d-ambitwistor models at $\scri$ allowing us to represent copies of~$Lw_{1+\infty}$ for both the self-dual and anti-self-dual sectors within the same model.

We begin in Section~\ref{Sec:SDwinf} with a brief review of $w_{1+\infty}$, its loop algebra and explain how it is realized in terms of twistor space and self-dual gravity. Section~\ref{Sec:TSigma} reviews the twistor sigma model, and its relationship to self-dual gravity at null infinity through the projection from asymptotic twistor space to $\scri$~\cite{Eastwood:1982}. We review how the model at degree-1 computes the MHV sector of tree-level graviton scattering.
In Section~\ref{subsec:Lw} we show how asymptotic symmetries are expressed in terms of the twistor sigma model; using the operator product expansion (OPE) of the model we show that these are controlled by the loop algebra $Lw_{1+\infty}$. Indeed, the twistor sigma model shows how the holomorphic curves of the non-linear graviton construction provide the most basic realization of this algebra.

Sections \ref{subsec:vert} and \ref{subsec:soft} explore the soft expansion of a positive helicity graviton in terms of vertex operators in the twistor sigma model. We show that this expansion gives the generators of $Lw_{1+\infty}$ and produces the infinite tower of soft graviton symmetries identified in~\cite{Guevara:2021abz,Strominger:2021lvk}.
Section~\ref{sec:ambi} outlines a generalization of these symmetries to both self-dual and anti-self-dual sectors of gravity by means of the 4d ambitwistor string~\cite{Geyer:2014fka} at $\scri$ \cite{Adamo:2019ipt, Geyer:2014lca}, pointing to avenues of future work. We conclude with some remarks regarding choices of~$(2,2)$ vs.~$(1,3)$ signature, quantization of the twistor sigma models, and their relation to the celestial holography programme.

\section[Lw\_\{1+infty\} and self-dual gravity]{$\boldsymbol{Lw_{1+\infty}}$ and self-dual gravity}\label{Sec:SDwinf}

The algebra $w_{1+\infty}$ arises as the Lie algebra of the Poisson structure (or area) preserving diffeomorphisms of the plane~\cite{Bakas:1989xu,Bakas:1989mz, Hoppe:1988gk}, although it can also be viewed as the classical limit of the~$W_{1+\infty}$ algebra associated to two-dimensional conformal field theories with higher-spin conserved currents~\cite{Fateev:1987zh,Fateev:1987vh, Zamolodchikov:1985wn}~-- see~\cite{Pope:1991ig} for a review. In this section we recall the basic structure of $w_{1+\infty}$, its loop algebra $Lw_{1+\infty}$, and their realization in twistor space through the non-linear graviton construction.

\subsection[Poisson diffeomorphisms and Lw\_\{1+infty\}]{Poisson diffeomorphisms and $\boldsymbol{Lw_{1+\infty}}$}

Let $\mu^{\dot\alpha}=\big(\mu^{\dot0},\mu^{\dot1}\big)$ be coordinates on the plane, with Poisson structure
\begin{gather}\label{Poisson-str}
\{ f,g\}:= \varepsilon^{\dot\alpha\dot\beta} \frac{\p f}{\p \mu^{\dot\alpha}} \frac{\p g}{\p \mu^{\dot\beta}} , \qquad \varepsilon^{\dot\alpha\dot\beta}=\varepsilon^{[\dot\alpha\dot\beta]}, \qquad\varepsilon^{\dot 0\dot 1}=1 .
\end{gather}
Elements of the Lie algebra of Poisson diffeomorphisms can be decomposed into polynomial Hamiltonians on the $\mu^{\dot\alpha}$-plane of degree $2p-2\in\Z_{\geq0}$:
\begin{gather*}
w^p_m:=\big(\mu^{\dot 0}\big)^{p+m-1} \big(\mu^{\dot 1}\big)^{p-m-1} , \qquad |m|\leq p-1 ,
\end{gather*}
so that $p\pm m-1 \in \Z_{\geq 0}$. The Poisson bracket acting on these elements gives
\begin{gather*}
\big\{w^p_{m},w^q_{n}\big\}=2  (m (q-1)-n (p-1) ) w^{p+q-2}_{m+n} .
\end{gather*}
This defines the commutation relations of the basis elements $w^p_m$ of $w_{1+\infty}$. Here, the `1' in $1+\infty$ refers to the central element of degree $2p-2=0$.

The loop algebra $Lw_{1+\infty}$ of $w_{1+\infty}$ can be represented by introducing a complex coordinate $\lambda\in\C$, where the loop is parametrized by $|\lambda|=1$. Alternatively, $\lambda$ can be viewed as an affine coordinate on the Riemann sphere $S^2\cong\CP^1$: if $\lambda_{\alpha}=(\lambda_0,\lambda_1)$ are homogeneous coordinates on~$\CP^1$, then on the patch where $\lambda_0\neq0$ we can identify $\lambda\equiv\lambda_1/\lambda_0$. With this, the generators of $Lw_{1+\infty}$ can be written as
\begin{gather}\label{Lwgens}
g^p_{m,r}:=\frac{w^p_m}{\lambda_0^{2p-4-r} \lambda_{1}^r}=\frac{w^p_m}{\lambda^r} ,
\end{gather}
where in the second equality we have chosen a scaling for the homogeneous coordinates in which $\lambda_0=1$. The Poisson bracket \eqref{Poisson-str} gives
\begin{gather}\label{Lw-inf-comm}
\big\{g^p_{m,r},g^q_{n,s}\big\}=2  (m (q-1)-n (p-1) ) g^{p+q-2}_{m+n,r+s} ,
\end{gather}
which define the Lie bracket of the loop algebra $Lw_{1+\infty}$.
Later we will also introduce the parameter $z$ on the equator of $\CP^1$ so that the~$g^p_{m,r}$ arise as the coefficients of the formal Laurent series appropriate to a contour around $\lambda=z$ in
\begin{gather}\label{g(z)}
g^p_m(z)=\frac{g^p_m}{\lambda-z}=\sum_{r\in\Z}g^p_{m,r} z^{r-1} ,
\end{gather}
defining a field insertion at the point $\lambda=z$.\footnote{This uses the language of 2d quantum fields, but this paper~-- excepting Section~\ref{sec:ambi}~-- mostly concerns the semi-classical limit.}

\subsection{Realization on twistor space}

The twistor space $\PT$ of complexified Minkowski space (i.e., $\C^4$ equipped with the holomorphic Minkowski metric) is an open subset of~$\CP^3$. If $Z^A=\big(\mu^{\dot\alpha},\lambda_{\alpha}\big)$ are four homogeneous coordinates on $\CP^3$, then twistor space is the open subset $\PT=\big\{Z\in\CP^3\,|\,\lambda_{\alpha}\neq0\big\}$. The relationship between $\PT$ and complexified Minkowski space is non-local: a point $x^{\alpha\dot\alpha}$ in the complexified space-time corresponds to a holomorphic, linearly embedded Riemann sphere in $\PT$ defined by $\mu^{\dot\alpha}=x^{\alpha\dot\alpha}\lambda_{\alpha}$.

Twistor space admits a natural fibration over $\CP^1$
\begin{gather}\label{fibration}
p\colon \ \PT\rightarrow\CP^1, \qquad p(Z)=\lambda_\alpha ,
\end{gather}
with $\lambda_{\alpha}$ serving as homogeneous coordinates on the Riemann sphere (this is possible precisely because $\lambda_{\alpha}\neq0$ on $\PT$). The fibres of $p$ are 2-planes~$\C^2$ with complex coordinates~$\mu^{\dal}$. Twistor space also admits the holomorphic Poisson structure~\eqref{Poisson-str}, where the Poisson bracket is trivially extended to act on functions that depend on $\lambda_{\alpha}$ as well as $\mu^{\dot\alpha}$. It provides a non-degenerate symplectic structure on every fibre.

One of the central results of twistor theory is the \emph{non-linear graviton} theorem:
\begin{Theorem}[Penrose~\cite{Penrose:1976js}]
There is a $1:1$ correspondence between self-dual Ricci-flat holomorphic metrics on regions in $\C^4$, and complex deformations $\CPT$ of twistor space $\PT$ that preserve the fibration $p\colon \CPT\rightarrow\CP^1$ and the Poisson structure~\eqref{Poisson-str} on the fibres of~$p$ defined on the neighbourhood of a line in $\CPT$ with normal bundle $\cO(1)\oplus\cO(1)$.
\end{Theorem}
Here, the holomorphic metrics on regions in $\C^4$ can be thought of as arising from complexification of an analytic split-signature or Riemannian self-dual 4-manifold, or as Newman's $\cH$-spaces defined by complexified self-dual characteristic data at null infinity~\cite{Ko:1981, Newman:1976gc}.

\begin{figure}[t]\centering
\includegraphics[scale=.80]{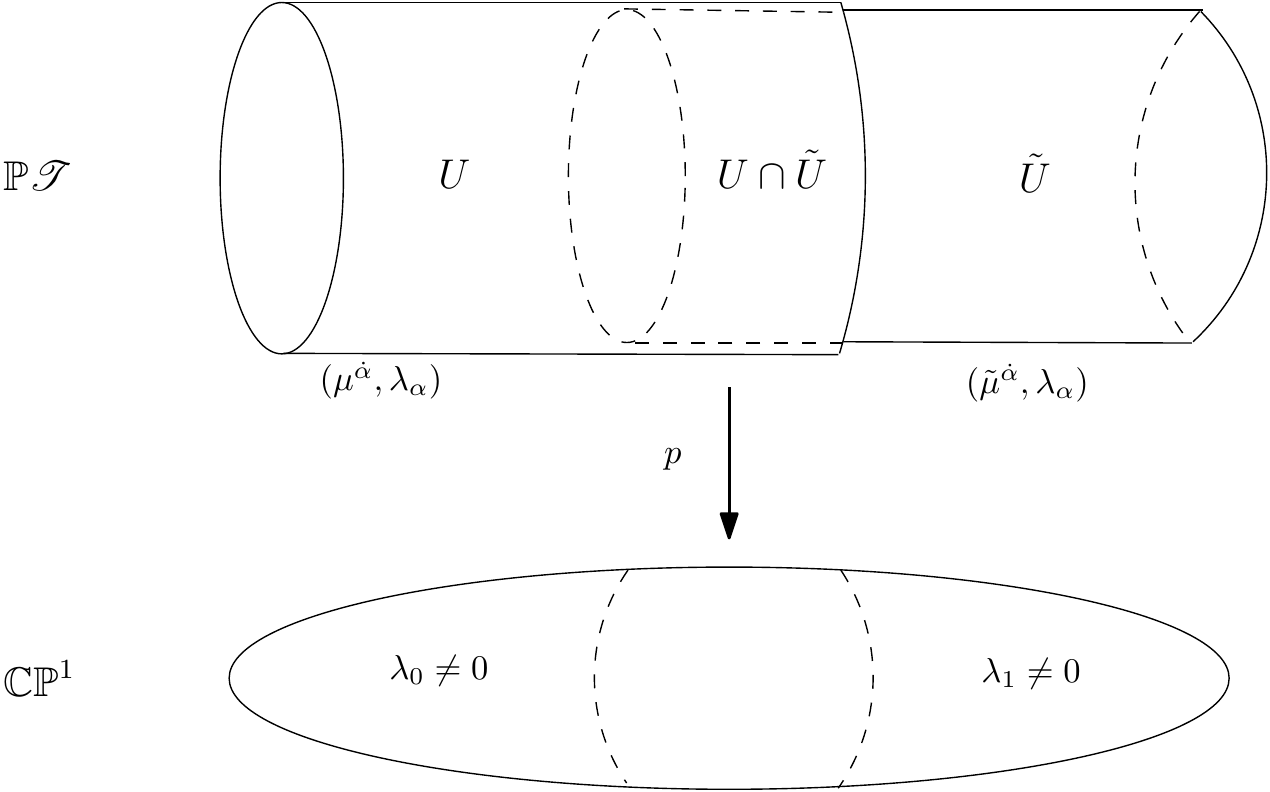}
\caption{The deformed twistor space $\CPT$ in terms of a patching fibred over $\CP^1$.}\label{Patching}
\end{figure}

In Penrose's original paper (see~\cite[Section 6]{Penrose:1976js}), the complex deformations of twistor space were described by deforming the patching functions of $\PT$ (thought of as a complex manifold) between the two coordinate patches
\begin{gather*}
U=\{ \lambda_0\neq 0\} , \qquad \tilde U=\{\lambda_1\neq 0\} ,
\end{gather*}
with coordinates $Z=\big(\mu^{\dot\alpha},\lambda_\alpha\big)$ and $\tilde Z=\big(\tilde{\mu}^{\dot\alpha},\lambda_\alpha\big)$, respectively. Since the deformations preserve the projection to the Riemann sphere, the coordinates $\lambda_{\alpha}$ on the two patches are identified on the overlap; see Figure~\ref{Patching}. In order to preserve the Poisson structure~\eqref{Poisson-str}, a generating function $G\big(\lambda_\alpha,\mu^{\dot 0},\tilde \mu^{\dot 1}\big)$ of homogeneity degree two is used to define the patching of the $\mu $-coordinates (implicitly) by
\begin{gather*}
\mu^{\dot 1}=\frac{\p G}{\p \mu^{\dot 0}} , \qquad \tilde \mu^{\dot 0}=\frac{\p G}{\p\tilde \mu^{\dot 1}} .
\end{gather*}
It is easy to see that this preserves the Poisson structure on any fibre of $\CPT\to\CP^1$, since~$G$ generates canonical transformations on the fibres.

Infinitesimally, deformations of such a twistor space are determined by Hamiltonians  $g(Z)=\delta G$ of homogeneity degree two. Such a $g$ should therefore be defined on the intersection $U\cap\tilde{U}$ of the coordinate patches, meaning that its expansion is polynomial in $\mu^{\dot\alpha}$ but Laurent in $\lambda=\lambda_1/\lambda_0$. These requirements mean that $g(Z)$ is expanded in the generators of the loop algebra $Lw_{1+\infty}$ given by \eqref{Lwgens}. In other words, $g^p_{m,r}$ form a basis of positive helicity (since deformations of the twistor space correspond to self-dual curvature in space-time) graviton states in linear theory, with the commutation relations~\eqref{Lw-inf-comm} thought of as the Lie algebra of the loop group of area preserving diffeomorphisms.\footnote{Here we do not address convergence issues; to make sense of such, one would need to consider a semi-group, etc.}

In linear theory, the wavefunctions corresponding to $g^p_{m,r}$ can be represented on space-time using standard integral formulae evaluated on twistor lines (cf.~\cite{Penrose:1969ae,Penrose:1986uia}):
\begin{gather}
\tilde{\psi}_{\dot\alpha_1 \ldots \dot\alpha_4}(x) =\oint \frac{\d\lambda}{2\pi\im}\left.\frac{\p^4 g^p_{m,r}}{\p\mu^{\dot\alpha_1}\cdots \p\mu^{\dot\alpha_4}} \right|_{\mu^{\dot\alpha}=x^{\alpha\dot\alpha}\lambda_{\alpha}} ,\nonumber 
\\
h_{\al\dal\beta\dot\beta}(x)=\iota_\alpha \iota_\beta \oint \frac{\d\lambda}{2\pi\im} \left.\frac{\p^2 g^p_{m,r}}{\p\mu^{\dot\alpha}\p\mu^{\dot\beta}} \right|_{\mu^{\dot\alpha}=x^{\alpha\dot\alpha}\lambda_{\alpha}} , \label{PTrans2}
\end{gather}
for the linearized self-dual Weyl spinor and metric perturbation respectively. Here, these formulae are written in the affine patch where $\lambda_0=1$. The contour integrals are taken around poles in the $\lambda$-plane and the constant spinor $\iota_{\alpha} = (0,1)$ is chosen so that $\la\iota \lambda\ra=\lambda_0=1$ on this affine patch. This choice of $\iota_{\alpha}$ amounts to a gauge fixing for the linear metric and drops out of the curvature. Clearly, these formulae give rise to polynomials in the space-time coordinates $x^{\alpha\dot\alpha}$
of degree $2p-6$ for the Weyl spinor or $2p-4$ for the metric. For example, with $g^{5/2}_{3/2, r}$ we find
\begin{gather*}
p=\frac{5}{2} ,\ m=\frac{3}{2}\colon \quad h_{\al\dal\beta\dot\beta}(x) = \iota_\al \iota_\beta \tilde o_{\dal} \tilde o_{\dot\beta} \big(x^{0\dot0} \delta_{r,1} + x^{1\dot0} \delta_{r,2}\big)
\end{gather*}
with $\tilde o_{\dal} = (1,0)$, etc. This is a mode of the sub-sub-leading soft graviton.

The solutions \eqref{PTrans2} directly yield modes of the conformally soft graviton wavefunctions of~\cite{Pasterski:2017kqt};
up to a constant multiple, these can be defined as the right hand side of
\begin{gather}\label{soft-modes}
\iota_\alpha \iota_\beta \oint \frac{\d\lambda}{2\pi\im} \left.\frac{\p^2g^p_{m,r}}{\p\mu^{\dot\alpha}\p\mu^{\dot\beta}}\right|_{\mu^{\dot\alpha}=x^{\alpha\dot\alpha}\lambda_{\alpha}}
 = \frac{\Gamma(p-m) \Gamma(p+m)}{(2\pi\im)^2 \Gamma(2p-3)}\oint\frac{\d z \d\tilde z}{z^r \tilde z^{p-m}} \iota_\al \iota_\beta \tilde z_{\dal} \tilde z_{\dot\beta} (q\cdot x)^{2p-4},
\end{gather}
where $z_\al=(1,z)$, $\tilde z_{\dal}=(1,\tilde z)$, $q_{\al\dal} = z_{\al} \tilde z_{\dal}$, and the contour on the right is a product of circles around $z=0$, and $\tilde z=0$; here $h_{\al\dal\beta\dot\beta}(x) = \iota_\al \iota_\beta \bar z_{\dal} \bar z_{\dot\beta} (q\cdot x)^{2p-4}/\Gamma(2p-3)$ is a generating series for these soft modes that will be defined more systematically later.

\section{Twistor sigma model and MHV amplitudes}\label{Sec:TSigma}

The non-linear graviton construction realizes the self-dual 4-manifold as the moduli space of degree one (rational) holomorphic curves in the deformed twistor space. In~\cite{Adamo:2021bej} we introduced a~sigma model for these holomorphic curves adapted to a Dolbeault description of the nonlinear graviton in which the complex structure is deformed by means of a global deformation of the d-bar operator, $\dbar\rightarrow \bar \nabla=\dbar + \cdots$, rather than the shift in the patching functions introduced in the previous section. In this language, our sigma model governs maps from the Riemann sphere to twistor space whose equation of motion determines the holomorphic twistor curves with respect to~$\bar \nabla$.

As shown in~\cite{Eastwood:1982}, such a Dolbeault description of the nonlinear graviton construction arises from an asymptotic twistor space defined by characteristic data at $\scri$. For curves of degree one, the solutions to the twistor sigma model yield the self-dual space-time; in this representation, the nonlinear graviton construction becomes a reformulation of Newman's $\mathcal{H}$-space construction~\cite{Newman:1976gc}. This connection with $\scri$ is what allows us to make contact with celestial holography. The MHV sector of tree-level graviton scattering arises at degree one, whereas for higher NMHV degree the boundary conditions of the model can be adapted to give rational curves of higher degree.

\subsection{Holomorphic curves and twistor sigma model}

While Penrose initially described complex deformations of twistor space in terms of patching functions, one can equivalently work with deformations of the almost complex structure that are integrable and preserve the fibration \eqref{fibration} as well as the Poisson structure \eqref{Poisson-str}. Such deformations are locally given by perturbing the Dolbeault operator,
\begin{gather}\label{c-str}
\nbar=\dbar+\varepsilon^{\dot\alpha\dot\beta} \frac{\partial\h}{\partial\mu^{\dot\alpha}} \frac{\partial}{\partial\mu^{\dot\beta}}=\dbar+\{\h,\;\} ,
\end{gather}
where $\dbar = \d\overline{Z^A} \p/\p\overline{Z^A}$ corresponds to the trivial complex structure on~$\PT$ for which $(\mu^{\dot\alpha},\lambda_{\alpha})$ are holomorphic, and $\h\in\Omega^{0,1}(\PT,\cO(2))$ with $(0,1)$-form components pointing along the~$\CP^1$ base of the fibration. In other words,
\begin{gather}\label{c-str2}
\h=h \D\bar{\lambda} , \qquad \D\bar{\lambda}\equiv[\bar{\lambda} \d\bar{\lambda}]=\bar{\lambda}^{\dot\alpha} \d\bar{\lambda}_{\dot\alpha} ,
\end{gather}
with $h$ a function on $\PT$ homogeneous of degree two in the holomorphic coordinates and $-2$ in the anti-holomorphic coordinates. It is straightforward to see that any almost complex structure of the form~\eqref{c-str} preserves the holomorphic fibration $\CPT\to\CP^1$ and Poisson structure. Integrability $\nbar^2=0$ is also immediate since $\D\bar{\lambda}\wedge\D\bar{\lambda}=0$. The linear perturbations associated to such deformations are obtained from the Penrose transforms,
\begin{gather}
\tilde{\psi}_{\dot\alpha_1 \ldots \dot\alpha_4}(x)  =\int_{\P^1}\D\lambda\wedge\left.\frac{\p^4 \h}{\p\mu^{\dot\alpha_1}\cdots \p\mu^{\dot\alpha_4}} \right|_{\mu^{\dot\alpha}=x^{\alpha\dot\alpha}\lambda_{\alpha}} , \label{DPTrans1}
\\
h_{\al\dal\beta\dot\beta}(x) =\int_{\P^1}\D\lambda\wedge\frac{\iota_\alpha \iota_\beta}{\la\iota \lambda\ra^2} \left.\frac{\p^2 \h}{\p\mu^{\dot\alpha}\p\mu^{\dot\beta}} \right|_{\mu^{\dot\alpha}=x^{\alpha\dot\alpha}\lambda_{\alpha}} ,\label{DPTrans2}
\end{gather}
where $\D\lambda = \lambda^\al \d\lambda_\al$. But we can also construct the fully non-linear self-dual vacuum metric associated to~$\h$ by employing the fact that such a metric is necessarily hyperk\"ahler.

A point in a self-dual vacuum space-time corresponds to a rational curve in $\CPT$ which is holomorphic with respect to the complex structure~\eqref{c-str}. Such a holomorphic curve can be described by viewing $\mu^{\dot\alpha}$ as a degree $-1$ map from $\CP^1$ to twistor space, with boundary conditions at the north and south poles of the Riemann sphere fixing all moduli of the curve. Letting $\sigma_{\ba}=(\sigma_0,\sigma_1)$ be homogeneous coordinates on~$\CP^1$, a degree one curve in twistor space is parametrized by
\begin{gather}\label{-1curve}
\lambda_{\alpha}(\sigma)=\left(\frac{1}{\sigma_0}, \frac{1}{\sigma_1}\right)=\frac{(1,\lambda)}{\sigma_0} , \qquad \mu^{\dot\alpha}(x,\sigma)=\frac{x^{\dot\alpha}}{\sigma_0}+\frac{\tilde{x}^{\dot\alpha}}{\sigma_1}+M^{\dot\alpha}(\sigma) .
\end{gather}
Here, the moduli of the curve have been fixed by specifying the pole structure in the first two terms of $\mu^{\dot\alpha}$ with $x^{\alpha\dot\alpha}=(x^{\dot\alpha},\tilde{x}^{\dot\alpha})$ providing coordinates on the self-dual space-time. The object~$M^{\dot\alpha}$ is smooth and homogeneous of weight $-1$ in $\sigma_\ba$; it is uniquely determined by the requirement that the curve is holomorphic with respect to \eqref{c-str}, i.e., that
\begin{gather}\label{hol-curve}
\dbar_\sigma M^{\dot\alpha}=\frac{\partial\h}{\partial\mu_{\dot\alpha}}(x,\sigma) .
\end{gather}
In other words, given the data $\h$ on $\CPT$ and the parametrization~\eqref{-1curve}, the self-dual space-time is reconstructed by solving~\eqref{hol-curve} for the holomorphic curves in twistor space.

In~\cite{Adamo:2021bej}, we showed that \eqref{hol-curve} arise as the Euler-Lagrange equations of a \emph{twistor sigma model}
\begin{gather}\label{tsm}
S[M]=\frac{1}{4\pi\im \hbar}\int_{\CP^1}\D\sigma\big([M \dbar_{\sigma}M]+2 \h(x,\sigma)\big) ,
\end{gather}
where $\D\sigma:=\sigma_{0}\d\sigma_1-\sigma_1\d\sigma_0$, $\big[M \dbar_\sigma M\big]:=\varepsilon_{\dot\alpha\dot\beta} M^{\dot\beta}\dbar_{\sigma}M^{\dot\alpha}$, and $\hbar$ is a formal parameter. Remarkably, this sigma model is directly related to the underlying self-dual geometry. Evaluating its on-shell action, it follows that (up to a constant)~\cite{Adamo:2021bej}
\begin{gather}\label{PlebProp}
\Omega(x)=\varepsilon_{\dot\alpha\dot\beta} x^{\dot\beta} \tilde{x}^{\dal} - \hbar S[M]\Big|_{\text{on-shell}} ,
\end{gather}
is the K\"ahler potential -- or first Pleba\'nski form~\cite{Plebanski:1975wn} -- for the self-dual metric. In particular, the metric is defined by the tetrad
\begin{gather*}
e^{\alpha\dot\alpha}=\big(\d x^{\dot\alpha}, \Omega^{\dot\alpha}{}_{\dot\beta} \d\tilde{x}^{\dot\beta}\big) , \qquad \Omega_{\dot\alpha\dot\beta}:=\frac{\partial^{2}\Omega}{\partial x^{\dot\alpha}\partial\tilde{x}^{\dot\beta}} ,
\end{gather*}
with self-duality corresponding to the `first heavenly equation' $\mathrm{det}(\Omega_{\dot\alpha\dot\beta})=2$.

\subsection[I and asymptotic twistor space]{$\boldsymbol{\scri}$ and asymptotic twistor space}

The non-linear graviton construction is directly related to the arena of celestial holography when the deformed twistor space is defined by the self-dual characteristic data at $\scri$.\footnote{By~$\scri$, we mean one of the future or past null conformal boundaries $\scri^{\pm}$; implicitly we will always choose the future boundary $\scri^+$ although it is trivial to work with $\scri^-$ instead.} From a twistor space $\CPT$, there is a natural projection
\begin{gather}\label{scriproj}
\CPT\rightarrow\scri_\C , \qquad \big(\mu^{\dot\alpha},\lambda_{\alpha}\big)\mapsto \big(u,\lambda_\al,\bar{\lambda}_{\dal}\big)=\big(\mu^{\dot\beta}\bar\lambda_{\dot\beta},\lambda_\al,\bar\lambda_{\dal}\big) ,
\end{gather}
where $\scri_\C\cong\C\times S^2$ is a partial complexification of the conformal boundary obtained by letting~$u$ become complex, but we do not complexify the $S^2$-factor. In particular, this identifies the $\CP^1$ base of the fibration~\eqref{fibration} with the celestial sphere~\cite{Eastwood:1982}.

Consider an asymptotically flat space-time with a Bondi--Sachs expansion that has been conformally rescaled by the conformal factor~$R^2$
with $R=r^{-1}$ and $r$ a standard radial coordinate to become
 \begin{gather*}
\d\hat{s}^2=-2 \d u \d R-\frac{4 \D\lambda \D\bar{\lambda}}{||\lambda||^4}+R \big(\bigma^0 \big(u,\lambda,\bar{\lambda}\big) \D\lambda^2+\bar{\bigma}^0\big(u,\lambda,\bar{\lambda}\big) \D\bar{\lambda}^2\big)+O\big(R^2\big) .
\end{gather*}
Here $||\lambda||^2= |\lambda_0|^2+|\lambda_1|^2$ yields a conformal factor for the round sphere in homogeneous coordinates $\lambda_\alpha$, and
 $\scri^+$ corresponds to $R\to 0$. The complex (spin- and conformal-weighted) function $\bigma^0$ encodes the asymptotic shear of the constant-$u$ hypersurfaces at $\scri$; this is the free characteristic data of the gravitational field (also often denoted by~$C_{zz}$). In a precise sense, $\bigma^0$~controls the anti-self-dual radiative degrees of freedom of the metric, with $\bar{\bigma}^0$ controlling the self-dual radiative degrees of freedom~\cite{Bondi:1962px,Newman:1961qr,Sachs:1961zz,Sachs:1962wk,Sachs:1962zzb}. The spin- and conformal-weights of~$\bigma^0$ dictate that it has the scaling property
\begin{gather}\label{shearscale}
\bigma^0\big(|b|^2 u, b\lambda, \bar{b}\bar{\lambda}\big)=\frac{\bar{b}}{b^3} \bigma^0\big(u,\lambda,\bar{\lambda}\big) , \qquad \bar{\bigma}^0\big(|b|^2 u, b\lambda, \bar{b}\bar{\lambda}\big)=\frac{b}{\bar{b}^3} \bar{\bigma}^0\big(u,\lambda,\bar{\lambda}\big) ,
\end{gather}
for any non-vanishing complex number $b$.

We define the complex structure as in \eqref{c-str}--\eqref{c-str2} on asymptotic twistor space by taking $\h=h\big(u,\lambda,\bar{\lambda}\big) \D\bar{\lambda}$ with $u$ given by the projection \eqref{scriproj} and
\begin{gather*}
h\big(u,\lambda,\bar{\lambda}\big)=\int^u \bar{\bigma}^0\big(s,\lambda,\bar{\lambda}\big) \d s .
\end{gather*}
The scaling property \eqref{shearscale} ensures that~$h$~-- and hence~$\h$~-- has the correct homogeneity on twistor space. Thus, the complex structure~\eqref{c-str} on $\CPT$ becomes
\begin{gather*}
\nbar=\dbar+\{\h,\;\}=\dbar+\bar{\lambda}_{\dot\alpha} \D\bar{\lambda} \bar{\bigma}^0\big(u,\lambda,\bar{\lambda}\big) \frac{\partial}{\partial\mu_{\dot\alpha}} .
\end{gather*}
Thus the deformed twistor space $\CPT$ is determined by the characteristic data. Such a twistor space is referred to as an \emph{asymptotic twistor space}; these twistor spaces can be characterised as those associated to Newman's $\cH$-spaces~\cite{Ko:1981, Newman:1976gc}, which are self-dual radiative space-times determined by complexified data with $\bigma^0=0$ but $\bar{\bigma}^0$ non-zero and independent of $\bigma^0$ on~$\scri_\C$, given by the~$\bar\bigma^0$ of the original Lorentzian space-time.

\subsection{From the sigma model to the MHV amplitude}

There is a direct connection between the twistor sigma model~\eqref{tsm} for asymptotic twistor spaces and the MHV helicity sector of tree-level graviton scattering. A tree-level gravitational MHV amplitude involves two negative helicity external gravitons and arbitrarily many positive helicity gravitons. When the total number of gravitons is $n$ (i.e., 2 negative helicity and $n-2$ positive helicity gravitons) there is a compact, elegant formula for this amplitude in a momentum eigenstate basis due to Hodges~\cite{Hodges:2012ym}:
\begin{gather}\label{HForm}
\cM_{n,0}=\delta^{4} \left(\sum_{i=1}^{n}k_i\right) \la1 2\ra^{8} \mathrm{det}^{\prime}(\HH) ,
\end{gather}
where overall factors of the gravitational coupling have been suppressed. In this expression, the $k_i^{\al\dal} = \kappa_i^\al \tilde\kappa_i^{\dal}$ are null momenta, gravitons 1 and 2 have been assigned negative helicity, $\HH$ is a~$(n-2)\times(n-2)$ matrix with entries
\begin{gather*}
\HH_{ij}=\frac{[i j]}{\la i j\ra} , \qquad i\neq j , \qquad \HH_{ii}=-\sum_{j\neq i}\frac{[i j]}{\la i j\ra} \frac{\la1 j\ra \la2 j\ra}{\la1 i\ra \la2 i\ra} ,
\end{gather*}
and the reduced determinant is defined by
\begin{gather*}
 \mathrm{det}^{\prime}(\HH):=\frac{|\HH^{i}_{i}|}{\la1 2\ra^{2} \la1 i\ra^{2} \la2 i\ra^{2}} .
\end{gather*}
It is easy to see that the choice of minor -- corresponding to a choice of one positive helicity external graviton~-- defining $\mathrm{det}^{\prime}(\HH)$ is arbitrary, so this formula nicely manifests the permutation symmetry of all positive helicity gravitons in the MHV scattering process.

Since the number of positive helicity gravitons in an MHV amplitude is arbitrary, it is natural to view them as being generated by the perturbative expansion of the two-point function of negative helicity gravitons on a non-linear self-dual background. Since the self-dual background in such a generating functional should be purely radiative (so that its perturbative limit produces positive helicity gravitons), its associated twistor space is an asymptotic twistor space.

This generating functional picture was first made precise in~\cite{Mason:2008jy} and later refined in~\cite{Adamo:2021bej}, with the result that the generating functional for MHV amplitudes can be written as
\begin{gather}\label{MHVgen}
-\la1 2\ra^{4} \int_{\cM}\d^{2}x \d^{2}\tilde{x} \e^{\im[x 1]+\im[\tilde{x} 2]} \Omega(x,\tilde{x})=\la1 2\ra^{4}\int_{\cM}\d^{2}x \d^{2}\tilde{x} \e^{\im[x 1]+\im[\tilde{x} 2]} S[M]\Big|_{\text{on-shell}} ,
\end{gather}
where $\cM$ is the self-dual background, $\Omega$ is its K\"ahler potential or first Pleba\'nski form and the equality follows thanks to \eqref{PlebProp}. Here, one implicitly adopts a 2-spinor basis in \eqref{-1curve} adapted to the momenta of the two negative helicity gravitons. This amounts to using $x^{\dal} = x^{\al\dal} \kappa_{1 \al}$ and $\tilde x^{\dal} = x^{\al\dal} \kappa_{2 \al}$ as coordinates on~$\cM$. We also set $\hbar=1$ for convenience; it will be reinstated when needed.

To view the self-dual background as a superposition of positive helicity gravitons, the complex structure of the asymptotic twistor space is taken to be
\begin{gather*}
\h(Z)=\sum_{i=3}^{n}\epsilon_i \h_{i}(Z;k_i) ,
\end{gather*}
where each $\h_i$ is a momentum eigenstate representative on twistor space:
\begin{gather}\label{mom-eig}
\h_i(Z;k_i)=\int_{\C^*}\frac{\d s_i}{s_i^3} \bar{\delta}^{2}(\kappa_{i \alpha}-s_i \lambda_{\alpha}) \e^{\im s_i[\mu i]} .
\end{gather}
Inserting this into the integral formulae \eqref{DPTrans1}--\eqref{DPTrans2}, one recovers the expected positive helicity momentum eigenstate on (complexified) Minkowski space:
\begin{gather*}
h_{i \alpha\dot\alpha\beta\dot\beta}(x)=\frac{\iota_{\alpha} \iota_{\beta} \tilde{\kappa}_{i \dot\alpha} \tilde{\kappa}_{i \dot\beta}}{\la \iota i\ra^2} \e^{\im k_i\cdot x} , \qquad \tilde{\psi}_{i \alpha\beta\gamma\delta}(x)=\tilde{\kappa}_{i \alpha} \tilde{\kappa}_{i \beta} \tilde{\kappa}_{i \gamma} \tilde{\kappa}_{i \delta} \e^{\im k_{i}\cdot x} .
\end{gather*}
Perturbatively expanding the generating functional~\eqref{MHVgen} then boils down to extracting the multi-linear piece of a tree-level correlation function involving insertions of these momentum eigenstates.

In particular, the on-shell action is evaluated using the tree-level, connected correlation functions of `vertex operators'
\begin{gather}\label{MHVcorr}
\left.\left(\prod_{i=3}^{n}\frac{\partial}{\partial\epsilon_i}\right) S[M]\Big|_{\text{on-shell}}\right|_{\epsilon_i=0}=\left\la \prod_{i=3}^{n}V_{i}\right\ra^{\mathrm{tree}}_{0} , \qquad V_i:=\int_{\CP^1}\D\sigma_i\wedge\h_{i}(Z(\sigma_i);k_i) ,
\end{gather}
in the two-dimensional CFT of the twistor sigma model with trivial complex structure. This means that the correlator is evaluated using the free OPE
\begin{gather}\label{SigOPE}
M^{\dot\alpha}(\sigma_i) M^{\dot\beta}(\sigma_j)\sim\frac{\varepsilon^{\dot\alpha\dot\beta}}{\sigma_i-\sigma_j} ,
\end{gather}
in the affine patch of $\CP^1$ where $\sigma_{\ba}=(1,\sigma)$. Here, the vertex operators are simply linear deformations of the sigma model action and the tree-level contribution is extracted from the generating functional for the connected correlator by taking $\hbar\to0$ as usual.

This computation is fairly straightforward as it involves keeping only single contractions in the OPE of any two vertex operators (see~\cite{Adamo:2021bej} for details). It gives
\begin{gather*}
 \left\la \prod_{i=3}^{n}V_{i}\right\ra^{\mathrm{tree}}_{0}=\frac{|\HH^{i}_{i}|}{\la1 i\ra^2 \la2 i\ra^2} \prod_{j=3}^{n}\e^{\im k_{i}\cdot x} ,
\end{gather*}
where the determinant arises as a result of the weighted matrix-tree theorem (which also ensures that the result is independent of the choice of $i$ singled out on the LHS) and all $\CP^1$ integrals can be performed against the delta functions appearing in~\eqref{mom-eig}. Feeding this into~\eqref{MHVgen} and using $\d^{2}x \d^{2}\tilde{x}=\la1 2\ra^2\d^{4}x$ immediately gives the Hodges formula~\eqref{HForm}, providing a first-principles derivation of tree-level MHV graviton scattering, which explains the appearance of `tree-summing' formulae~\cite{Bern:1998sv,Nguyen:2009jk} and the matrix-tree theorem~\cite{Adamo:2012xe, Feng:2012sy} in earlier literature.

By adapting the boundary conditions for the $\mu^{\dot\alpha}(\sigma)$ map, it is possible to formulate a higher-degree version of the twistor sigma model (i.e., by imposing boundary conditions at $d+1$ points on $\CP^1$). These higher degree models are related to other helicity sectors of the tree-level graviton $S$-matrix, with degree $d$ corresponding to N$^{d-1}$MHV amplitudes, although the generating functionals for $d>1$ cannot be derived directly from general relativity and require additional ingredients (albeit quite minimally) beyond the on-shell action of the twistor sigma model~\cite{Adamo:2021bej}.\looseness=-1

\section[From twistorial to celestial Lw\_\{1+infty\}]{From twistorial to celestial $\boldsymbol{Lw_{1+\infty}}$}\label{Sec:OPE}

With the self-dual sector of gravity on space-time captured by the twistor sigma model \eqref{tsm}, it is now straightforward to describe infinitesimal deformations and hence the symmetry algebra associated to the self-dual sector. Using the semi-classical OPE on the Riemann sphere defined by the sigma model, we first show how this produces the expected $Lw_{1+\infty}$ algebra. We go on to explain the relationship between graviton vertex operators and $Lw_{1+\infty}$ symmetry generators as a realization of a \v{C}ech--Dolbeault isomorphism within the model. We then give the soft expansion of these vertex operators/symmetry generators so as to yield the basis we introduced in Section~\ref{Sec:SDwinf}. Furthermore, using the relationship between the twistor sigma model and tree-level MHV scattering, we prove that this explicitly generates the action of celestial $Lw_{1+\infty}$ on positive helicity hard gravitons of~\cite{Guevara:2021abz,Strominger:2021lvk}.

\subsection[Lw\_\{1+infty\} charges and algebra]{$\boldsymbol{Lw_{1+\infty}}$ charges and algebra}\label{subsec:Lw}

The form of the complex structure \eqref{c-str}--\eqref{c-str2} on twistor space admits coordinate symmetries generated by Hamiltonians with respect to the Poisson structure \eqref{Poisson-str}. Such Hamiltonians $g\big(\mu^{\dot\alpha},\lam,\bar\lambda\big)$ must have homogeneity degree~2 in~$Z^{A}$ and be holomorphic\footnote{In fact, the requirement of holomorphicity in~$\mu^{\dot\alpha}$ can be dropped if more general~$\h$ are allowed (e.g.,~\cite{Mason:2007ct}).} in $\mu^{\dot\alpha}$ but not necessarily in $\lambda_{\alpha}$. The symmetry action is given by
\begin{gather}\label{mutrans}
\delta \mu^{\dot\alpha}= \big\{ g,\mu^{\dot\alpha}\big\}= \varepsilon^{\dot\beta\dal}\frac{ \p g}{\p \mu^{\dot\beta}} , \qquad \delta \h= \dbar g + \{\h,g\} ,
\end{gather}
which leads to a symmetry of the twistor sigma model action \eqref{tsm} when $\delta \h=0$, i.e., when $g$ satisfies $\delta\h= \dbar g + \{\h,g\}=0$ so that $g$ is holomorphic with respect to the deformed complex structure $\nbar$. For such $g$, Noether's theorem leads to the conserved charge{\samepage
\begin{gather}\label{can-charges}
Q_g= \oint g \D\sigma ,
\end{gather}
in the theory on $\CP^1$ defined by the sigma model.}

The OPE \eqref{SigOPE} extends from the `non-zero-mode' $M^{\dot\alpha}$ to the full twistor coordinate $\mu^{\dot\alpha}$ in the obvious way (since the two differ only by zero modes):
\begin{gather}\label{SigOPE2}
\mu^{\dot \alpha}(\sigma) \mu^{\dot \beta}(\sigma')\sim \frac{\varepsilon^{\dot\alpha\dot\beta}}{\sigma-\sigma'} ,
\end{gather}
on the usual affine patch where $\sigma_{\ba}=(1,\sigma)$. This in turn induces a semi-classical OPE for the Hamiltonian functions $g$ given by the Poisson bracket
\begin{gather}
g(Z(\sigma)) g'(Z(\sigma'))\sim \frac{1}{\sigma-\sigma'} \{g, g'\}(\sigma') , \label{can-OPE}
\end{gather}
with higher order singularities being neglected at tree-level in the sigma model. Thus, the OPE encodes the loop algebra of the Poisson diffeomorphisms of the $\mu^{\dot\alpha}$-plane with loop variable $\lambda$. The charges $Q_g$ given by~\eqref{can-charges} generate canonical transformations of the $\mu^{\dot\alpha}$-plane with canonical commutation relations
\begin{gather*}
[ Q_g, Q_{g'}]=Q_{\{g,g'\}} ,
\end{gather*}
also arising from the semi-classical OPE.

Poisson diffeomorphisms generated by Hamiltonians satisfying $\delta\h=\dbar g + \{\h,g\}=0$ do not deform the space-time K\"ahler scalar \eqref{PlebProp} as they leave the on-shell action of the twistor sigma model invariant. As a result, the functions $g$ must generically have singularities in $\lambda$ to encode non-trivial symmetry transformations of the self-dual sector. Consider a BMS supertranslation corresponding to $\delta u= f(\lambda,\bar{\lambda})$ where $f$ has homogeneity $+1$ in the homogeneous coordinates~$\lambda_{\alpha}$,~$\bar{\lambda}_{\dot\alpha}$ of the celestial sphere. Using the projection~\eqref{scriproj} from asymptotic twistor space to~$\scri_\C$, this corresponds to a transformation
\begin{gather*}
\delta\mu^{\dot\alpha}=\frac{\p f}{\p \bar\lambda_{\dot\alpha}} ,
\end{gather*}
which is in turn generated by the Hamiltonian
\begin{gather}\label{ST}
g_{\mathrm{ST}}= \left[\mu \frac{\p f}{\p\bar\lambda}\right] ,
\end{gather}
under \eqref{mutrans}. When $f\big(\lambda,\bar\lambda\big)=a^{\alpha\dot\alpha}\lambda_\alpha\bar\lambda_{\dot\alpha}$, these are just the usual translations. Similarly, self-dual/dotted Lorentz super-rotations (of the extended BMS algebra~\cite{Barnich:2009se}) are generated by
\begin{gather}\label{SR}
g_{\mathrm{SR}}= \tilde{L}_{\dot\alpha\dot\beta}(\lambda,\bar\lambda) \mu^{\dot\alpha} \mu^{\dot\beta} ,
\end{gather}
where $\tilde{L}_{\dot\alpha\dot\beta}$ is homogeneous of degree zero in $\lambda_{\alpha}$, $\bar{\lambda}_{\dot\alpha}$. When $\tilde{L}_{\dot\alpha\dot\beta}$ depends only on $\bar{\lambda}_{\dot\alpha}$, this reduces to a standard Lorentz rotation.

In general, the transformations \eqref{ST}, \eqref{SR} are \emph{not} symmetries of the sigma model action, since $\delta\h\neq0$. Indeed, for the charge \eqref{can-charges} to be conserved, one requires $g$ to be holomorphic on twistor space; on a flat background (i.e., $\h=0$) this requires $g$ to be globally-defined and one simply obtains the Poincar\'e algebra. A generic supertranslation \eqref{ST} or superrotation \eqref{SR} will have poles in $\lambda$, so to go beyond the Poincar\'e group~-- or on any curved background -- one must consider Hamiltonians $g$ which have singularities in a local holomorphic coordinate system. Such singularities indicate that these functions change the gravitational data: they are no longer simply symmetries.

Thus, generic charges \eqref{can-charges} generate canonical transformations of the $\mu^{\dot\alpha}$-plane that depend on~$\lambda$. Given the overall homogeneity constraint on~$g$~-- namely, that it is homogeneous of degree~$2$ on twistor space -- each Hamiltonian function can be decomposed into modes $g^{p}_{m,r}$ of the form~\eqref{Lwgens}. The OPE~\eqref{can-OPE} then dictates that these modes have Poisson brackets
\begin{gather*}
\big\{g^p_{m,r},g^q_{n,s}\big\}=2  (m (q-1)-n (p-1) ) g^{p+q-2}_{m+n,r+s} ,
\end{gather*}
which are precisely the commutation relations of $Lw_{1+\infty}$ given previously in~\eqref{Lw-inf-comm}. These can be expressed in terms of the semiclassical OPE of the operators \eqref{g(z)} as
\begin{gather*}
\big\{g^p_{m}(z),g^q_{n}(z')\big\}=2 \frac{m (q-1)-n (p-1)}{z-z'} \big(g^{p+q-2}_{m+n}(z)-g^{p+q-2}_{m+n}(z')\big) ,
\end{gather*}
Thus, the structure of the twistor sigma model naturally encodes $Lw_{1+\infty}$ in terms of its infinitesimal deformations.

\subsection{Vertex operators and currents and soft limits}\label{subsec:vert}

The relationship between vertex operators in the sigma model and $Lw_{1+\infty}$ currents relies on the \v{C}ech--Dolbeault correspondence. While Penrose's original formulation of the non-linear graviton construction utilized patching functions for the deformed twistor space, the twistor sigma model works directly with the deformed Dolbeault operator for the complex structure. In this Dolbeault approach, $\h\in H^1(\PT,\cO(2))$ is represented by the $(0,1)$-form $\h\in\Omega^{0,1}(\PT,\cO(2))$ obeying \mbox{$\dbar\h=0$}; for asymptotic twistor space with $\h=h(u,\lambda,\bar{\lambda}) \D\bar{\lambda}$ these conditions are automatic.

To find the \v{C}ech representative corresponding to such an~$\h$, locally on an open subset $U_a$ of twistor space, $\dbar\h=0$ can be solved by $\h=\dbar g_a$ for some smooth function $g_a$ of homogeneity~$2$. The differences $g_{ab}:=g_a-g_b$ are therefore holomorphic functions on $U_{ab}:=U_a\cap U_b$, defined up to the addition of holomorphic functions that extend over the~$U_a$; such $g_{ab}$ equivalence classes provide \v{C}ech representatives of $\h$ (with the open-set indices $a,b,\dots$ usually suppressed).

Our key example is the momentum eigenstate~\eqref{mom-eig}. Here we now separate out the frequency~$\omega$ explicitly so that we can also expand in $\omega$ to give the Taylor series around $\omega =0$ which then define the leading and subleading soft limits of graviton insertions. Thus, taking for simplicity an outgoing graviton, we write
\begin{gather*}
k^{\alpha\dot\alpha}=\omega z^\alpha \bar z^{\dot\alpha} , \qquad z_\alpha=(1,z) , \qquad \bar z_{\dot\alpha}=(1,\bar z) ,
\end{gather*}
so that for example $\kappa_\alpha=\sqrt{\omega} z_\alpha$, $\tilde\kappa_{\dal} = \sqrt{\omega} \bar z_{\dal}$ are the standard spinor helicity variables.
The Dolbeault representative is given by
simply re-writing~\eqref{mom-eig} to account for the frequency:
\begin{gather*}
\h=\la\iota \lambda\ra^3 \bar{\delta}(\la\lambda z\ra) \e^{\im \omega \frac{[\mu \bar{z}]}{\la\iota \lambda\ra}} , \qquad \bar{\delta}(\lambda)=\frac{1}{2\pi\im} \dbar\left(\frac{1}{\lambda}\right) ,
\end{gather*}
where $\iota_{\alpha}=(0,1)$ is a constant spinor basis element.
The corresponding \v{C}ech representative is
\begin{gather}\label{Cechrep}
g=\frac{\la\iota \lambda\ra^3}{2\pi\im} \frac{1}{\la\lambda z\ra} \e^{\im \omega \frac{[\mu \bar z]}{\la\iota \lambda\ra}} ,
\end{gather}
with the choice of $\iota_{\alpha}$ now reflecting the \v Cech cohomology gauge freedom. The relevant open sets are given by covering the Riemann sphere with $U_0$ containing $\la\lambda \iota\ra = 0$ and $U_1$ containing $\la\lambda z\ra=0$; the overlap is a neighbourhood of the contour $\gamma_z$
\begin{gather*}
\gamma^\epsilon_z=\left\{\left|\frac{\la\lambda z\ra}{\la\lambda \iota\ra}\right|=\epsilon\right\} ,
\end{gather*}
for some small $\epsilon>0$. Inside of $\gamma_z^\epsilon$, the vertex operator for $\h$ obeys{\samepage
\begin{gather}\label{CechDol}
V_{\h}=\int_{\CP^1}\h\wedge\D\sigma=\oint_{\gamma_z}g \D\sigma =Q_{g} ,
\end{gather}
by Cauchy's theorem.}

In the soft limit as $\omega\rightarrow 0$, the exponential factor in~\eqref{Cechrep} can be expanded in powers of~$\omega$ to obtain combinations of the $Lw_{1+\infty}$ generators $g^{p}_{m}(z)$ as coefficients of~$\omega^{2p-2}$ (taking for simplicity the affine patch where $\la\iota \lambda\ra=1$ and $\la\lambda z\ra=\lambda-z$). For $2p-2=1,2$, this gives the standard correspondence between the leading and sub-leading soft graviton theorems and generators of supertranslations and superrotations, respectively; for $2p-2\geq3$ we obtain an infinite tower of soft graviton symmetries corresponding to higher-order generators of~$Lw_{1+\infty}$.

We can also make precise contact with the incarnation of $Lw_{1+\infty}$ first noted in the context of celestial holography by~\cite{Strominger:2021lvk}. Consider a positive helicty graviton boost eigenstate of conformal weight $\Delta$ inserted at the point $z_\al = (1,z)$, $\bar z_{\dal} = (1,\bar z)$ on the celestial sphere.\footnote{This will become the celestial torus in split signature when $\bar z$ is independent of $z$.} Its Dolbeault twistor representative reads~\cite{Adamo:2019ipt}
\begin{gather*}
\h = \frac{(-\im \eps)^{-\Delta} \Gamma(\Delta-2)}{[\mu \bar z]^{\Delta-2}} \bar\delta_\Delta(\la\lambda z\ra) ,
\end{gather*}
where $\eps=\pm1$ denotes whether it is outgoing or incoming, and we have defined a holomorphic delta function of weight $\Delta$ in $\lambda_\al$:
\begin{gather*}
\bar\delta_\Delta(\la\lambda z\ra) := \la\iota \lambda\ra^{\Delta+1} \bar\delta(\la\lambda z\ra) .
\end{gather*}
Again, $\iota_\al=(0,1)$ so that $\la\iota \lambda\ra=\lambda_0$, $\la\iota z\ra=1$, etc. Inserting this in the Penrose integral formula~\eqref{DPTrans2}, one finds the expected wavefunction of a spin 2 positive helicity boost eigenstate:
\begin{gather*}
h_{\al\dal\beta\dot\beta}(x) = (-\im \eps)^{-\Delta} \Gamma(\Delta) \frac{\iota_\al \iota_\beta \bar z_{\dal} \bar z_{\dot\beta}}{(q\cdot x)^\Delta} ,
\end{gather*}
with $q_{\al\dal} = z_\al \bar z_{\dal}$. This is gauge equivalent to a spin 2 conformal primary graviton~\cite{Pasterski:2017kqt} whose modes we considered in~\eqref{soft-modes}.

Without loss of generality, we focus on outgoing particles for which $\eps=+1$. Conformally soft gravitons are obtained by taking residues at $\Delta = k =2,1,0,-1,\dots$:
\begin{gather}\label{hksoft}
\h^k_\text{soft} = \text{Res}_{\Delta=k}\h = \frac{\im^{-k}}{(2-k)!} [\mu \bar z]^{2-k} \bar\delta_k(\la\lambda z\ra) .
\end{gather}
Substituting $[\mu \bar z] = \mu^{\dot0}+\bar z \mu^{\dot1}$ in~\eqref{hksoft}, it can be binomially expanded into a polynomial in $\bar z$ to get $3-k$ holomorphic currents. In doing this, we use the index relabeling $k=4-2p$. Hence,
\begin{gather}\label{hpsoft}
\h^{4-2p}_\text{soft} = \im^{2p-4} \bar\delta_{4-2p}(\la\lambda z\ra)\sum_{m=1-p}^{p-1}\frac{\bar z^{p-1-m} w^p_m}{(p-m-1)! (p+m-1)!} ,
\end{gather}
where
\begin{gather*}
w^p_m  = (\mu^{\dot0})^{p+m-1}(\mu^{\dot1})^{p-m-1} ,\qquad p=1,\frac{3}{2},2,\frac{5}{2},\dots .
\end{gather*}
Remarkably, the combinatorial rescaling by $(p-m-1)! (p+m-1)!$ that was crucial for the identification of $w_{1+\infty}$ in~\cite{Strominger:2021lvk} emerges naturally here via twistor space. The modes in~\eqref{hpsoft} give Dolbeault twistor representatives
\begin{gather*}
\im^{2p-4} \bar\delta_{4-2p}(\la\lambda z\ra) w^p_m
\end{gather*}
for the various soft gravitons that are in correspondence with celestial $Lw_{1+\infty}$ generators. Thus, as explained in~\eqref{CechDol}, in the twistor sigma model these correspond to charges
\begin{gather}\label{Qpm}
Q^p_m(z)=\frac{\im^{2p}}{2\pi\im}\oint_{\gamma_z} g^p_m(z) \d\sigma= \frac{\im^{2p}}{2\pi\im}\oint_{\gamma_z} \frac{w^{p}_{m}(\sigma)}{\la\lambda(\sigma) z\ra} \la\iota \lambda(\sigma)\ra^{5-2p} \D\sigma ,
\end{gather}
with the contour integral taken around the pole at $\la\lambda z\ra=\lambda-z=0$ (the second equality is a re-writing in homogeneous coordinates of the first). These are the $w_{1+\infty}$ currents generating Poisson diffeomorphisms on the $\lambda = z$ fibre of twistor space.

\subsection{Soft graviton symmetries}\label{subsec:soft}

Finally, we show that the twistorial action of $w_{1+\infty}$ on hard gravitons is equivalent to the celestial action of $w_{1+\infty}$ given in~\cite{Guevara:2021abz,Himwich:2021dau,Jiang:2021ovh,Strominger:2021lvk}. More precisely, the OPE between positive helicity soft and hard graviton vertex operators in the twistor sigma model maps to the celestial OPE between the conformally soft gravitons and hard gravitons (as dictated by collinear limits or asymptotic symmetries). We also show that the action of a~$w_{1+\infty}$ generator on a negative helicity graviton gives rise to the mixed helicity soft-hard celestial OPE, but leave the interpretation of this at the level of the sigma model correlators to future work.

{\bf Action on positive helicity gravitons.} Let $\h_{\Delta_i}(\sigma_i)$ be the twistor representative of an outgoing, positive helicity graviton with conformal dimension $\Delta_i$ and celestial positions $(z_i,\bar{z}_i)$:
\begin{gather*}
\h_{\Delta_i}(\sigma_i) = \frac{\im^{\Delta_i} \Gamma(\Delta_i-2)}{[\mu(\sigma_i) \bar z_i]^{\Delta_i-2}} \bar\delta_{\Delta_i}(\la\lambda(\sigma_i) z_i\ra) ,
\end{gather*}
where $z_{i \al}\equiv(1,z_i)$, $\bar z_{i \dal}\equiv(1,\bar z_i)$ as usual. We label this representative with $\Delta_i$ and suppress~$z_i$,~$\bar z_i$ for brevity. Acting on it with the soft charge~$Q^p_m$ in~\eqref{Qpm}, and using the sigma model OPE~\eqref{SigOPE2}, we find
\begin{align*}
Q^{p}_{m}(z) \h_{\Delta_i}(\sigma_i)&\sim \frac{\im^{2p}}{2\pi\im}\oint \frac{\la\iota \lambda(\sigma)\ra^{5-2p}}{\la\lambda(\sigma) z\ra} \frac{\partial w^{p}_{m}}{\partial\mu_{\dot\alpha}}(\sigma) \frac{\partial\h_{\Delta_i}}{\partial\mu^{\dot\alpha}}(\sigma_i) \frac{\D\sigma}{\sigma-\sigma_i} \\
& = -\im^{2p} \frac{\la\iota \lambda(\sigma_i)\ra^{5-2p}}{\la\lambda(\sigma_i) z\ra} \frac{\partial w^{p}_{m}}{\partial\mu_{\dot\alpha}}(\sigma_i) \frac{\partial\h_{\Delta_i}}{\partial\mu^{\dot\alpha}}(\sigma_i) ,
\end{align*}
where the contour integral has been evaluated by deforming\footnote{Although there are potentially other poles that might be picked up by this deformation, these are either subleading in the celestial OPE limit~$z-z_i\to 0$ or do not contribute to tree-level correlators in the twistor sigma model.} the contour from the $\la\lambda(\sigma) z\ra=0$ pole to the pole at $\sigma=\sigma_i$. As usual, we have only kept a single contraction in the OPE as we want to insert this in tree correlators at the end.

On the support of the holomorphic delta function $\bar{\delta}_{\Delta_i}(\la\lambda(\sigma_i) z_i\ra)$ appearing in $\h_{\Delta_i}$, the action of the soft charge can be further simplified to
\begin{align}
Q^{p}_{m}(z) \h_{\Delta_i}(\sigma_i)&\sim-\frac{\im^{2p}}{\la z_i z\ra} \la\iota \lambda(\sigma_i)\ra^{4-2p} \frac{\partial w^{p}_{m}}{\partial\mu_{\dot\alpha}}(\sigma_i) \frac{\partial\h_{\Delta_i}}{\partial\mu^{\dot\alpha}}(\sigma_i) \nonumber\\
 & =\frac{\im^{2p}}{z-z_i} \frac{\{w_m^p,\h_{\Delta_i}\}(\sigma_i)}{\la\iota \lambda(\sigma_i)\ra^{2p-4}} .\label{tope}
\end{align}
Thus, the OPE between a soft graviton current and a conformal primary hard graviton is given by the action of~$Lw_{1+\infty}$ in its canonical (in the sense of the Poisson bracket) representation. As expected, this fact is most directly visible on twistor space.

We can now prove that the \emph{celestial} action of a soft graviton symmetry on a positive helicity hard graviton arises from the Poisson bracket in~\eqref{tope}. Using $[\mu \bar z_i] = \mu^{\dot0}+\bar z_i \mu^{\dot1}$, it follows that
\begin{gather}
\frac{\{w_m^p,\h_{\Delta_i}\}(\sigma_i)}{\la\iota \lambda(\sigma_i)\ra^{2p-4}} = -\Big[(p+m-1) \bar z_i \big(\mu^{\dot0}\big)^{p+m-2}(\sigma_i) \big(\mu^{\dot1}\big)^{p-m-1}(\sigma_i)   \nonumber\\
\qquad{} - (p-m-1) \big(\mu^{\dot0}\big)^{p+m-1}(\sigma_i) \big(\mu^{\dot1}\big)^{p-m-2}(\sigma_i)\Big] \frac{\im^{\Delta_i} \Gamma(\Delta_i-1)}{[\mu(\sigma_i) \bar z_i]^{\Delta_i-1}} \bar\delta_{\Delta_i-2p+4}(\la\lambda(\sigma_i) z_i\ra) .\label{wmph}
\end{gather}
Next, we have the intertwining relations
\begin{gather*}
\mu^{\dot1} \frac{\Gamma(a)}{[\mu \bar z_i]^{a}} = -\dbar_i \frac{\Gamma(a-1)}{[\mu \bar z_i]^{a-1}} ,\qquad\mu^{\dot0}\frac{\Gamma(a)}{[\mu \bar z_i]^{a}} = (\bar z_i\dbar_i+a-1) \frac{\Gamma(a-1)}{[\mu \bar z_i]^{a-1}} ,
\end{gather*}
where $\dbar_i\equiv\p/\p\bar z_i$ and $a\neq1$. Applying these iteratively to the right hand side of \eqref{wmph}, one can re-express the OPE \eqref{tope} as
\begin{gather}
Q^p_m(z) \h_{\Delta_i}(\sigma_i) \sim \frac{(-1)^{p+m}}{z-z_i}\left[(p+m-1) \bar z_i \biggl(\prod_{r=1}^{p+m-2}\big(\bar z_i\dbar_i+\Delta_i-1-r\big)\biggr) \dbar_i^{p-m-1} \right.\nonumber\\
\left.\hphantom{Q^p_m(z) \h_{\Delta_i}(\sigma_i) \sim}{}
+ (p-m-1) \biggl(\prod_{r=1}^{p+m-1}\big(\bar z_i\dbar_i+\Delta_i-1-r\big)\biggr) \dbar_i^{p-m-2}\right]\h_{\Delta_i-2p+4}(\sigma_i) .\!\!\!\label{cope}
\end{gather}
Expanding the bracketed operators gives the celestial OPE
\begin{gather}
Q^p_m(z) \h_{\Delta_i}(\sigma_i)\sim \frac{(-1)^{p+m}}{z-z_i}\sum_{\ell=0}^{p+m-1}{p+m-1\choose\ell}\frac{(2p-2-\ell) \Gamma(\Delta_i-1)}{\Gamma(\Delta_i-1-\ell)}\nonumber\\
\hphantom{Q^p_m(z) \h_{\Delta_i}(\sigma_i)\sim}{}
\times\bar z_i^{p+m-1-\ell}\dbar_i^{2p-3-\ell}\h_{\Delta_i-2p+4}(\sigma_i) ,\label{positive-action}
\end{gather}
previously found in the literature~\cite{Himwich:2021dau,Jiang:2021ovh}. Inserting these relations into the sigma model tree correlators \eqref{MHVcorr} straightforwardly produces the corresponding celestial OPE between a $w_{1+\infty}$ current and a hard graviton. This gives rise to the tower of conformally soft theorems and asymptotic symmetries found in \cite{Guevara:2019ypd,Guevara:2021abz,Strominger:2021lvk}.
For instance, one can easily verify the actions of supertranslation, superrotation as well as the sub-sub-leading soft graviton symmetries.

Notice how the twistor description produces the celestial OPE in a factorized form \eqref{cope} which is highly non-trivial to see in a direct calculation of Mellin-transformed amplitudes in the collinear limit. It is this factorized form that hides the representation theory of $w_{1+\infty}$ and makes contact with its symplectic origins.

{\bf Action on negative helicity gravitons.} In the twistor sigma model, negative helicity gravitons are not represented by vertex operators, but classically one can still define a twistor representative for a negative helicity graviton. It is given by a $(0,1)$-form of weight $-6$ in $Z$: $\tilde \h\in\Omega^{0,1}(\PT,\cO(-6))$. It generates a graviton on space-time with purely negative helicity curvature computed by the Penrose transform
\begin{gather*}
\psi_{\al\beta\gamma\delta}(x)=\int_{\P^1}\D\lambda\wedge\lambda_\al \lambda_\beta \lambda_\gamma \lambda_\delta\left.\tilde\h \right|_{\mu^{\dot\alpha}=x^{\alpha\dot\alpha}\lambda_{\alpha}} .
\end{gather*}
We are free to associate to this an operator $\tilde\h(Z(\sigma))$ in our sigma model. For instance, with the~$i^\text{th}$ outgoing negative helicity graviton boost eigenstate we associate the operator
\begin{gather*}
\tilde\h_{\Delta_i}(\sigma_i) = \frac{\im^{\Delta_i} \Gamma(\Delta_i+2)}{[\mu(\sigma_i) \bar z_i]^{\Delta_i+2}} \bar\delta_{\Delta_i-4}(\la\lambda(\sigma_i) z_i\ra) .
\end{gather*}
The corresponding classical twistor representative can be checked to produce the space-time curvature of a negative helicity boost eigenstate.

Repeating the derivation of \eqref{tope} yields the sigma model OPE
\begin{gather*}
Q^{p}_{m}(z) \tilde\h_{\Delta_i}(\sigma_i) \sim \frac{\im^{2p}}{z-z_i} \frac{\big\{w_m^p,\tilde\h_{\Delta_i}\big\}(\sigma_i)}{\la\iota \lambda(\sigma_i)\ra^{2p-4}} .
\end{gather*}
Computing the Poisson bracket then gives
\begin{gather*}
Q^p_m(z) \tilde\h_{\Delta_i}(\sigma_i) \sim \frac{(-1)^{p+m}}{z-z_i}\left[(p+m-1) \bar z_i \biggl(\prod_{r=1}^{p+m-2}\big(\bar z_i\dbar_i+\Delta_i+3-r\big)\biggr) \dbar_i^{p-m-1} \right.\\
\left.
\hphantom{Q^p_m(z) \tilde\h_{\Delta_i}(\sigma_i) \sim}{}
+ (p-m-1) \biggl(\prod_{r=1}^{p+m-1}(\bar z_i\dbar_i+\Delta_i+3-r)\biggr) \dbar_i^{p-m-2}\right]\tilde\h_{\Delta_i-2p+4}(\sigma_i) .
\end{gather*}
Expanding the derivative operators again produces the result in the literature~\cite{Himwich:2021dau,Jiang:2021ovh},
\begin{gather}
Q^p_m(z) \tilde\h_{\Delta_i}(\sigma_i)\sim \frac{(-1)^{p+m}}{z-z_i}\sum_{\ell=0}^{p+m-1}{p+m-1\choose\ell}\frac{(2p-2-\ell) \Gamma(\Delta_i+3)}{\Gamma(\Delta_i+3-\ell)}\nonumber\\
\hphantom{Q^p_m(z) \tilde\h_{\Delta_i}(\sigma_i)\sim}{}
\times\bar z_i^{p+m-1-\ell}\dbar_i^{2p-3-\ell}\tilde\h_{\Delta_i-2p+4}(\sigma_i) .\label{negative-action}
\end{gather}
Although we can no longer simply insert this in sigma model correlators to claim that this computes the soft-hard collinear limit, it is nevertheless remarkable that the twistorial action of~$w_{1+\infty}$ on negative helicity gravitons still maps to the corresponding celestial action.

\section{The lift to 4d ambitwistor string}\label{sec:ambi}

The twistor sigma model \eqref{tsm} is intrinsically chiral; while it can be used to define generating functionals for the full tree-level $S$-matrix of gravity beyond the MHV helicity sector, this requires additional ingredients which are inserted by hand~\cite{Adamo:2021bej}. A consequence of this chirality is that we find only the copy of $Lw_{1+\infty}$ associated with the self-dual/positive helicity soft sector; of course, there should be another copy associated with the anti-self-dual/negative helicity soft sector. Here, we observe that both copies of $Lw_{1+\infty}$ are naturally found in the four-dimensional ambitwistor string~\cite{Geyer:2014fka}, a CFT on the Riemann sphere whose correlation functions generate the tree-level $S$-matrix of gravity. We remark that the correlation functions in the 4d ambitwistor strings are now fully quantum, unlike the computations in the twistor sigma model \eqref{tsm} which are all semi-classical. Nevertheless, they faithfully represent only the semi-classical $Lw_{1+\infty}$.

Although we do not display the computations here, an identical calculation for the gravitational twistor string~\cite{Skinner:2013xp} yields a representation of $Lw_{1+\infty}$ as described here in the 4d ambitwistor string. However, it does not obviously have an anti-self-dual $\widetilde{Lw}_{1+\infty}$ sector and so may be a~better vehicle for seeing the action of the self-dual $Lw_{1+\infty}$ on the whole amplitude (i.e., all helicity sectors). However, the action of $\widetilde{Lw}_{1+\infty}$ is no longer manifest and will not be realized locally. This parity asymmetry is a familiar feature of twistor strings (cf.~\cite{Witten:2004cp}).

\subsection{Lifting to ambitwistor space}

One can extend beyond the self-dual sector by lifting to \emph{ambitwistor space} $\A$ defined by
\begin{gather*}
\A=\big\{ \big(Z^A,\tilde Z_A\big)\in \C^4\times \big(\C^4\big)^*\,\big|\, Z\cdot \tilde Z=0\big\}\big/ \big\{ \big(Z,\tilde Z\big)\sim \big(bZ,b^{-1}\tilde Z\big), b\in\C^{*}\big\} .
\end{gather*}
This is the cotangent bundle of both projective twistor space and projective dual twistor space, $\A=T^*\PT=T^*\PT^*$ and so has a symplectic structure, with dual Poisson structure defined by
\begin{gather*}
\omega=\d\theta , \qquad \theta:=Z\cdot \d\tilde Z-\tilde Z\cdot \d Z , \qquad \{\,,\,\}_\A:=\frac{\p}{\p Z^A}\wedge \frac{\p}{\p \tilde Z_A} .
\end{gather*}
This structure does not break left-right symmetry, and deformations of $\PT$ and $\PT^*$ both determine deformations of~$\A$~\cite{Mason:1985sva,Baston:1987av}.

In particular, any vector field $V^A \p/\p Z^A$ on $\PT$ has a Hamiltonian lift to~$\A$ with Hamiltonian~$V^A\tilde{Z}_A$. This enables a lift of deformation Hamiltonians on $\PT$ and $\PT^*$ to give the ambitdextrous Hamiltonian~\cite{Baston:1987av}
\begin{gather*}
H_{g,\tilde g}=\tilde\lambda^{\dot\alpha} \frac{\p g}{\p\mu^{\dot\alpha}}+\lambda_\alpha \frac{\p \tilde g}{\p\tilde\mu_\alpha} , \qquad g\in H^1(\PT,\cO(2)),\qquad \tilde g\in H^1(\PT^*,\cO(2)) ,
\end{gather*}
where here $g$, $\tilde g$ are taken to be \v{C}ech representatives. The corresponding Hamiltonian vector field on $\A$ determines deformations of the complex structure on $\A$ that have self-dual part~$H^+_g$ determined by~$g(Z)$, and anti-self-dual part $H^-_{\tilde g}$ determined by $\tilde g\big(\tilde{Z}\big)$.

It is easy to see that with these Hamiltonian lifts, the Poisson bracket on ambitwistor space restricted to the self-dual sector reproduces the Poisson bracket \eqref{Poisson-str} on twistor space
\begin{gather*}
\big\{H^+_g,H^+_{g'}\big\}_\A= H^+_{\{g,g'\}_\PT} .
\end{gather*}
This then gives a lift of the $Lw_{1+\infty}$ action to $\A$. The $H^-_{\tilde g} $ similarly lift to give the anti-self-dual $\widetilde{Lw}_{1+\infty}$-action on $\A$. One can then consider the commutator of the self-dual and anti-self-dual parts:
\begin{gather}
\big\{ H^+_g,H^-_{\tilde g}\big\}_\A= \left\{ \left[\tilde \lambda \frac{\p g}{\p\mu}\right], \left\langle \lambda \frac{\p\tilde{g}}{\p\tilde \mu}\right\rangle \right\}_\A=\left[\tilde \lambda \frac{\p}{\p\mu}\right] \frac{\p g}{\p Z^A} \left\langle \lambda \frac{\p}{\p\tilde \mu}\right\rangle \frac{\p\tilde g}{\p \tilde Z_A} .\label{H2-contr}
\end{gather}
In terms of deformation theory, the right hand side defines a class in $H^2(\A,\cO(1,1))$ that obstructs the exponentiation of the deformation generated by $H_{g,\tilde g}$. However, this cohomology group vanishes for elementary reasons~\cite{Baston:1987av}, so the deformation determined by $H_{g,\tilde g}$ can indeed be exponentiated.\footnote{The formula for the obstruction \eqref{H2-contr} naturally extends to $\PT\times \PT^*$ where it does not generically vanish. It was shown in~\cite{Baston:1987av} that to leading order around $\A$ it gives on space-time the Eastwood--Dighton conformal invariant defined in terms of SD and ASD Weyl spinors by $\psi_{\alpha\beta\gamma\delta}\nabla^{\delta\dot\delta}\tilde \psi_{\dot\alpha\dot\beta\dot\gamma\dot\delta}- \psi \leftrightarrow \tilde \psi$. There it was interpreted as an obstruction to extending the curved version of $\A$ into a curved analogue of $\PT\times \PT^*$ in a gravitational version of~\cite{Isenberg:1978kk,Witten:1978xx}; this was later proved in the fully nonlinear regime~\cite{LeBrun:1991jh}.}

\subsection{The 4d ambitwistor string}

For our purposes, the four-dimensional ambitwistor string~\cite{Geyer:2016nsh, Geyer:2014fka,Geyer:2014lca} for gravity has bosonic target space fields\footnote{To realize space-time supersymmetry $\big(Z^A,\tilde Z_A\big)$ are extended to include fermionic coordinates \cite{Geyer:2014fka}.} $\big(Z^A,\tilde Z_A\big)$ that are spinors on the worldsheet with an ambitwistor analogue of worldsheet supersymmetry giving spinor-valued partners $\big(\rho^A,\tilde \rho_A\big)$ of opposite statistics and worldsheet action
\begin{gather*}
S=\int_\Sigma \im Z\cdot\dbar \tilde Z -\im \tilde Z \cdot \dbar Z+ \rho\cdot \dbar \tilde \rho +\tilde \rho \cdot \dbar \rho + S_{\mathrm{Ghosts}} ,
\end{gather*}
where $\Sigma\cong\CP^1$. Here, all symmetries of the worldsheet theory (including those generated by the ambitwistor current $Z\cdot\tilde{Z}$ and those generating worldsheet supersymmetry) are assumed to have been gauge-fixed, leading to ghost fields with action $S_{\mathrm{Ghosts}}$ and a corresponding BRST operator~$Q$ (see \cite[Section~5.3]{Geyer:2016nsh} for details).

The upshot of this BRST quantization is that a non-trivial correlator needs one vertex operator each of the form
\begin{gather*}
U_h=\int_\Sigma \delta^2(\nu) \h , \qquad \tilde U_{\tilde h}=\int_\Sigma \delta^2(\tilde \nu) \tilde \h .
\end{gather*}
Here the $\nu$ and $\tilde \nu$ are two-component, weightless, bosonic ghost fields whose zero-modes are fixed by integration directly against these delta functions. Descent yields the remaining vertex operators for a correlator as
\begin{gather*}
V_\h:=\int_\Sigma \big[\tilde \lambda \p_\mu\big] \h(Z) + L\cdot \p^2_{\mu}\h(Z) , \qquad \tilde V_{\tilde \h}:=\int_\Sigma \langle \lambda \p_{\tilde \mu }\rangle \tilde \h\big(\tilde Z\big) + \tilde L\cdot \p^2_{\tilde \mu} \tilde \h\big(\tilde Z\big) ,
\end{gather*}
where $L^{\dot\alpha\dot\beta}=\rho^{(\dot\alpha}\tilde\rho^{\dot\beta)}$ is a self-dual Lorentz current algebra and $\tilde L^{\alpha\beta}=\rho^{(\alpha}\tilde \rho^{\beta)}$ is an anti-self-dual Lorentz current algebra, both constructed from the $\rho$-$\tilde\rho$ fermion system. As before, we can use a \v Cech representation of the cohomology groups $H^1(\PT,\cO(2))$ and $H^1(\PT^*,\cO(2))$ to re-express the vertex operators in terms of currents as
\begin{gather*}
V_g=\oint_\gamma \big[\tilde \lambda \p_\mu\big] g(Z) + L\cdot \p^2_{\mu} g(Z) , \qquad \tilde V_{\tilde g}:=\oint_\gamma \langle \lambda \p_{\tilde \mu }\rangle \tilde g\big(\tilde Z\big) + \tilde L\cdot \p^2_{\tilde \mu} \tilde g
\big(\tilde Z\big),\end{gather*}
where $\gamma$ is a path in $\Sigma$ that separates the singular regions of both~$g$ and~$\tilde g$.

When the vertex operators are both self-dual a direct calculation shows that they simply represent the Poisson bracket \eqref{Poisson-str} on $\PT$:
\begin{gather*}
V_g V_{g'}\sim \frac{1}{\sigma-\sigma'} V_{\{g,g'\}_\PT} +\cdots .
\end{gather*}
Hence, by expanding $g$ in the modes~\eqref{Lwgens} this gives $Lw_{1+\infty}$; an identical statement on dual twistor space gives $\widetilde{Lw}_{1+\infty}$ for the anti-self-dual vertex operators. However, when one vertex operator is self-dual and the other anti-self-dual, we have
\begin{gather*}
V_g\cdot \tilde V_{\tilde g} \sim \frac{1}{\sigma-\sigma'} \Big( \big\{\big[\tilde \lambda \p_\mu\big] g(Z) , \langle \lambda \p_{\tilde \mu }\rangle \tilde g\big(\tilde Z\big) \big\}_\A+\cdots \Big),
\end{gather*}
where the displayed term is the first of the semi-classical contribution as in \eqref{H2-contr} but now the $+\cdots$ contain infinitely many singular contributions with arbitrarily many contractions. In the computation of the full correlation function~\cite{Geyer:2016nsh, Geyer:2014fka}, these contributions are summed for momentum eigenstates by taking them into the Lagrangian in the path integral to produce the polarized or refined scattering equations. Remarkably, it is possible to show that this OPE encodes collinear splitting in a momentum eigenstate basis, or celestial OPEs in a conformal primary basis, although the mixed helicity case is particularly subtle~\cite{Adamo:2021tba}.

\section{Discussion}\label{Sec:Disc}

We have seen that the $Lw_{1+\infty}$ recently discovered in the soft OPE obtained from celestial amplitudes~\cite{Strominger:2021lvk} has a local representation as Poisson diffeomorphisms of the fibres of asymptotic twistor space and has its origin in Penrose's nonlinear graviton construction~\cite{Penrose:1976js}. The $Lw_{1+\infty}$ algebra associated with the positive helicity soft sector arises directly from a twistor sigma model describing self-dual gravity~\cite{Adamo:2021bej}, and this is explicitly identified with the soft OPE algebra on the celestial sphere. This acts on both the self-dual and anti-self-dual parts of the complexified gravitational data via a local action on twistor space when both parts are expressed as cohomology classes on that space via~\eqref{positive-action} and~\eqref{negative-action}. It is possible to obtain the helicity conjugate copy $\widetilde{Lw}_{1+\infty}$ of $Lw_{1+\infty}$ via its natural local representation on the conjugate (or dual) twistor space. One can see them both acting together by lifting to ambitwistor space, and to recover the correct celestial OPE one must use the fully quantum worldsheet CFT of the ambitwistor string~\cite{Adamo:2021tba}. There are many open questions and future directions related to the work in this paper; we conclude by touching on a few of them.

{\bf Split versus Lorentzian signature.} In this paper, we worked with the complexification of~$Lw_{1+\infty}$ realized as the holomorphic Poisson diffeomorphisms of $\C^2$. For polynomial generators of $w_{1+\infty}$ there are no analytic continuation issues. Viewing this $\C^2$ as the fibres of asymptotic twistor space, this complexification of $Lw_{1+\infty}$ corresponds to a partial complexification of null infinity $\scri\rightarrow\scri_\C\cong\C\times S^2$, by \eqref{scriproj}. Such a partial complexification is intrinsically associated with an underlying Lorentzian-real space-time, since the space of null directions remains the celestial sphere.

\looseness=1 Conversely, the real version of $Lw_{1+\infty}$ is not appropriate for Lorentzian signature data. In the real-valued case, $Lw_{1+\infty}$ gives the Poisson diffeomorphisms of $\R^2$ so the twistor components $\mu^{\dot\alpha}$ are themselves taken to be real-valued. Such a real-valued twistor space is appropriate to split signature space-time, where the celestial sphere is replaced by a celestial torus. The assumption of split signature is often used in celestial holography to disentangle the self- and anti-self-dual sectors and expedite various integral transformations (cf.~\cite{Atanasov:2021oyu,Atanasov:2021cje,Guevara:2021tvr,Guevara:2021abz,Sharma:2021gcz,Strominger:2021lvk}). In that context, the combinatorial factors and re-labelings appearing in the expansion~\eqref{hpsoft} emerge from a light transform, while in this paper we saw that this was not necessary.

\looseness=1
The split-signature versions of the twistor constructions used here have realizations in terms of holomorphic discs~\cite{Lebrun:2007, Mason:2005qu}, suggesting an `open string' approach to the subject in split signature. Similarly, light transforms in celestial CFT are related to half-Fourier transforms to real twistor space~\cite{Sharma:2021gcz}. Thus, although we have here been able to retain physical Lorentz signature, an explicit split signature version of the constructions in this paper might well be interesting.

{\bf Yang--Mills and Einstein--Yang--Mills.} Gauge theory also contains an infinite tower of conformally soft gluons, associated to conformal weights $\Delta=1,0,-1,\ldots$ in a conformal primary basis, and these have an associated infinite-dimensional current-like symmetry algebra not unlike $Lw_{1+\infty}$~\cite{Strominger:2021lvk}. Twistor theory also admits an elegant description of self-dual Yang--Mills theory via the Ward correspondence~\cite{Ward:1977ta}, the gauge theory analogue of the non-linear graviton. One can build a gauge theory version of the twistor sigma model which operationalizes the Ward correspondence; following similar steps to those presented here will yield a twistorial representation of the infinite-dimensional algebra associated with the positive helicity soft gluon sector. This algebra can be seen as arising from the natural action of gauge transformations on the twistor data for self-dual Yang--Mills on asymptotic twistor space.
Similar statements are possible for Einstein--Yang--Mills and the action of soft graviton symmetries on gluons.
However, the ambitwistor string provides a more direct root to studying soft gluons and celestial OPEs in pure Yang--Mills (for which there is a consistent worldsheet model) and even for Einstein--Yang--Mills, where a fully consistent worldsheet theory is not known~\cite{Adamo:2021tba}.

{\bf Towards quantization.} The twistor sigma model \eqref{tsm} gives rise to gravitational amplitudes via its classical action and the corresponding tree expansion; by contrast twistor strings or ambitwistor strings produce amplitudes as fully quantum correlations functions in the worldsheet CFT. This distinction leaves room for one to ask what the twistor sigma model could correspond to if treated quantum mechanically. In particular, there is scope for this to give rise to some theory of self-dual quantum gravity, for instance as envisaged by~\cite{Ooguri:1990ww,Ooguri:1991fp} for the $\cN=2$ string. For instance, the `quantum' (i.e., finite $\hbar$) MHV graviton amplitude produced by the twistor sigma model can be computed~\cite{Adamo:2021bej}:
\begin{gather*}
\cM_n=
\delta^{4} \left(\sum_{r=1}^{n}k_r\right) \la1 2\ra^{2n} \prod_{i=3}^{n}\frac{1}{\la1 i\ra^2 \la2 i\ra^2} \exp\left[-\frac{\im \hbar}{8\pi}\sum_{j\neq i}\frac{[i j]}{\la i j\ra} \frac{\la1 i\ra^2 \la2 j\ra^2}{\la 1 2\ra^2}\right] ,
\end{gather*}
although its physical properties and interpretation remain to be explored. It would also be intriguing to make contact with the $*$-algebra definition of the quantum $W_{1+\infty}$-algebra as described in~\cite{Pope:1991ig} and the Moyal deformations of the Poisson structure associated to self-dual gravity proposed by~\cite{Strachan:1992em} which are closely connected also to Penrose's Palatial twistor ideas~\cite{Penrose:2015lla}. It would be interesting to track the twistor-theoretic component of the other appearances of $W$-infinity algebras in the literature.

\subsection*{Acknowledgements}

We are grateful to Andrew Strominger for discussions. TA is supported by a Royal Society University Research Fellowship and by the Leverhulme Trust (RPG-2020-386). LJM is supported in part by the STFC grant ST/T000864/1. AS is supported by a Mathematical Institute Studentship, Oxford and by the ERC grant GALOP ID:~724638. We would also like to thank the SIGMA team for their courage and dedication, working through difficult circumstances in Kyiv in the face of the Russian invasion and aggression.

\pdfbookmark[1]{References}{ref}
\LastPageEnding

\end{document}